\documentclass[amsmath,amssymb,superscriptaddress,nobalancelastpage,prl,twocolumn,showpacs]{revtex4}
\usepackage{graphicx}
\usepackage{color}

\begin{document}


\title{Nernst and Seebeck Coefficients of the Cuprate Superconductor YBa$_2$Cu$_3$O$_{6.67}$:
A Study of Fermi Surface Reconstruction}


\author{J. Chang}
\affiliation{D\'epartement de physique
\& RQMP, Universit\'e de Sherbrooke, Sherbrooke, Canada}

\author{R. Daou}
\altaffiliation{Current address: Max Planck Institute for Chemical Physics of Solids, 01187 Dresden, Germany}
\affiliation{D\'epartement de physique
\& RQMP, Universit\'e de Sherbrooke, Sherbrooke, Canada}

\author{Cyril Proust}
\affiliation{Laboratoire National des Champs Magn\'etiques Intenses 
(CNRS), 31432 Toulouse, France}
\affiliation{Canadian Institute for Advanced Research, Toronto, Canada}

\author{David LeBoeuf}
\affiliation{D\'epartement de physique
\& RQMP, Universit\'e de Sherbrooke, Sherbrooke, Canada}

\author{Nicolas Doiron-Leyraud} 
\affiliation{D\'epartement de physique
\& RQMP, Universit\'e de Sherbrooke, Sherbrooke, Canada}

\author{Francis Lalibert\'e} 
\affiliation{D\'epartement de physique
\& RQMP, Universit\'e de Sherbrooke, Sherbrooke, Canada}

\author{B. Pingault} 
\affiliation{D\'epartement de physique
\& RQMP, Universit\'e de Sherbrooke, Sherbrooke, Canada}

\author{B.~J. Ramshaw} 
\affiliation{Department of Physics \& Astronomy,
University of British Columbia, Vancouver, Canada}

\author{Ruixing Liang} 
\affiliation{Department of Physics \& Astronomy,
University of British Columbia, Vancouver, Canada}
\affiliation{Canadian Institute for Advanced Research, Toronto, Canada}

\author{D.~A. Bonn} 
\affiliation{Department of Physics \& Astronomy,
University of British Columbia, Vancouver, Canada}
\affiliation{Canadian Institute for Advanced Research, Toronto, Canada}

\author{W.~N. Hardy} 
\affiliation{Department of Physics \& Astronomy,
University of British Columbia, Vancouver, Canada}
\affiliation{Canadian Institute for Advanced Research, Toronto, Canada}

\author{H. Takagi}
\affiliation{Department of Advanced Materials, University of Tokyo, Kashiwa 277-8561, Japan}

\author{A. B. Antunes}
\affiliation{Laboratoire National des Champs Magn\'etiques Intenses
(CNRS), 38042 Grenoble, France}

\author{I. Sheikin}
\affiliation{Laboratoire National des Champs Magn\'etiques Intenses 
 (CNRS), 38042 Grenoble, France}

\author{K. Behnia}
\affiliation{LPEM (UPMC-CNRS), ESPCI, 75231 Paris, France}

\author{Louis Taillefer}
\altaffiliation{E-mail: louis.taillefer@physique.usherbrooke.ca.}
\affiliation{D\'epartement de physique 
\& RQMP, Universit\'e de Sherbrooke, Sherbrooke, Canada} 
\affiliation{Canadian Institute for Advanced Research, Toronto, Canada}

\date{\today}


\begin{abstract}
The Seebeck and Nernst coefficients $S$ and $\nu$ of the cuprate 
superconductor YBa$_2$Cu$_3$O$_y$ (YBCO) were measured in a single crystal 
with doping $p = 0.12$ in magnetic fields up to $H = 28$~T.
Down to $T=9$~K, $\nu$ becomes independent of field by $H \simeq 30$~T, 
showing that superconducting fluctuations have become negligible.
In this field-induced normal state, $S/T$ and $\nu/T$ are both large and 
negative in the $T \to 0$ limit, with the magnitude and sign of $S/T$ 
consistent with the small electron-like Fermi surface pocket
detected previously by quantum oscillations and the Hall effect.
The change of sign in $S(T)$ at $T \simeq 50$~K is remarkably similar to 
that observed in La$_{2-x}$Ba$_x$CuO$_4$, La$_{2-x-y}$Nd$_y$Sr$_x$CuO$_4$ 
and La$_{2-x-y}$Eu$_y$Sr$_x$CuO$_4$, where it is clearly associated with 
the onset of stripe order. We propose that a similar density-wave 
mechanism causes the Fermi surface reconstruction in YBCO.

\end{abstract}


\maketitle


A major hurdle in understanding high-temperature superconductivity is the nature of the pseudogap phase.  
No consensus has yet been 
reached on whether this enigmatic phase is a precursor of superconductivity or a second ordered phase \cite{Norman2005}. 
One way to shed light on this question is to study the ground state of the pseudogap phase in the absence of superconductivity, 
achieved by applying a strong enough magnetic field. 
This approach has recently revealed a qualitative change in the Fermi surface of cuprates measured via quantum oscillations,
from a large hole-like cylinder in the overdoped regime outside the pseudogap 
phase \cite{Vignolle2008} to a 
Fermi surface containing small electron-like pockets~\cite{LeBoeuf2007} in the underdoped regime inside the pseudogap phase \cite{LeBoeuf2007,Doiron-Leyraud2007,Yelland2008,Bangura2008,Jaudet2008}.
Because the presence of an electron pocket in the Fermi surface of hole-doped cuprates almost certainly implies that the lattice translational symmetry
is broken by some density-wave order \cite{Chakravarty2008}, it is important to confirm the electron-like nature of the Fermi pocket detected in YBa$_2$Cu$_3$O$_y$ (YBCO), 
and elucidate the mechanism that causes it to emerge.

In this Letter, we show that: 
1) the low-temperature Nernst coefficient of YBCO at $p = 0.12$ is independent of field by $H \simeq 30$~T, proof that the vortex contribution is negligible by then, 
and the normal state has been reached;
2) the magnitude and negative sign of the thermopower at low temperature are consistent with the frequency and cyclotron mass of quantum oscillations
only if these come from orbits around an electron pocket. 
From the fact that both the thermopower and the Hall coefficient of YBCO are very similar to 
those of three cuprate materials exhibiting `stripe' order, 
a form of spin/charge density wave, we infer that the Fermi surface of underdoped YBCO also
undergoes a reconstruction due to a similar form of spin and/or charge ordering.


\begin{figure}
\begin{center}
\includegraphics[width=0.44\textwidth]{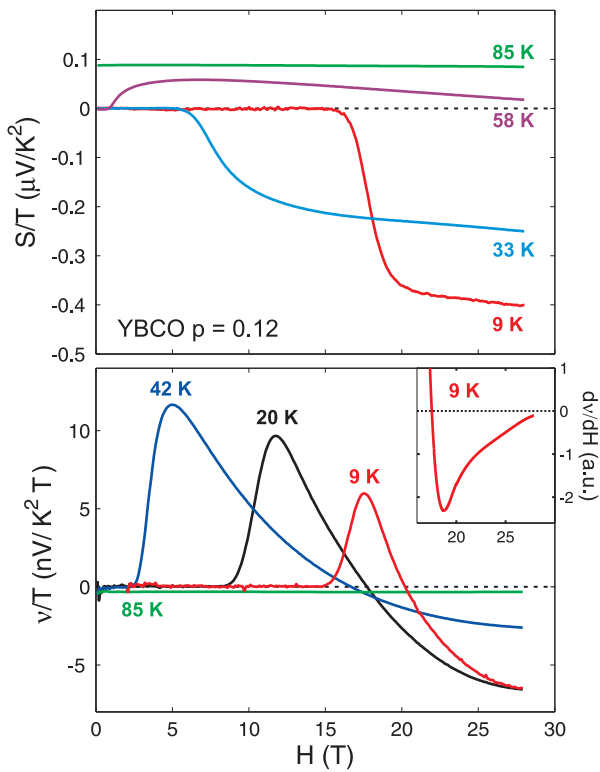}
\caption{
Thermo-electric coefficients of YBCO at $p = 0.12$ as a function of magnetic field $H$.
Upper panel:
Seebeck coefficient $S$ plotted as $S/T$ vs $H$, for temperatures as indicated.
Lower panel:
Nernst coefficient $\nu$ plotted as $\nu/T$ vs $H$, for temperatures as indicated.
{\it Inset:} Derivative of the 9-K isotherm (in arbitrary units), showing that $d\nu/dH \to 0$ as $H \to 30$~T.
}
\label{fig:fig1}
\end{center}
\end{figure}


When a temperature difference $\Delta T$ is applied along the $x$-axis of a metallic sample, a longitudinal voltage $V_x$ develops
across a length $L$, and the Seebeck coefficient (or thermopower) is defined as $S \equiv V_x / \Delta T$. 
In the presence of a perpendicular magnetic field $H$ (along the $z$-axis), a transverse voltage $V_y$ also develops across the width $w$ of the sample, 
and the Nernst coefficient is defined as $\nu \equiv (V_y / \Delta T) (1/H) (L/w)$. 
An estimate of the magnitude of these coefficients in metals can be obtained from the following simple expressions, valid as $T \to 0$ \cite{Behnia2004,Behnia2009}:
\begin{equation}
\frac{S}{T} \simeq \pm \frac{\pi^2}{2} \frac{k_{\rm B}}{e} \frac{1}{T_{\rm F}}
\end{equation} 
\begin{equation}
\frac{\nu}{T} \simeq \frac{\pi^2}{3} \frac{k_{\rm B}}{e} \frac{\mu}{T_{\rm F}}
\end{equation} 
where $k_{\rm B}$ is Boltzmann's constant, $e$ is the electron charge, $\mu$ is the carrier mobility and $T_{\rm F}$ the Fermi temperature.    
The sign of $S$ is controlled by the carrier type: positive for holes, negative for electrons.
$\nu$ can be of either sign, with no direct relation to carrier type, but the contribution of moving vortices is always positive \cite{Behnia2009}.
Both expressions have been found to work well in the $T=0$ limit for a wide range of metals \cite{Behnia2004,Behnia2009}.


\begin{figure}
\begin{center}
\includegraphics[width=0.44\textwidth]{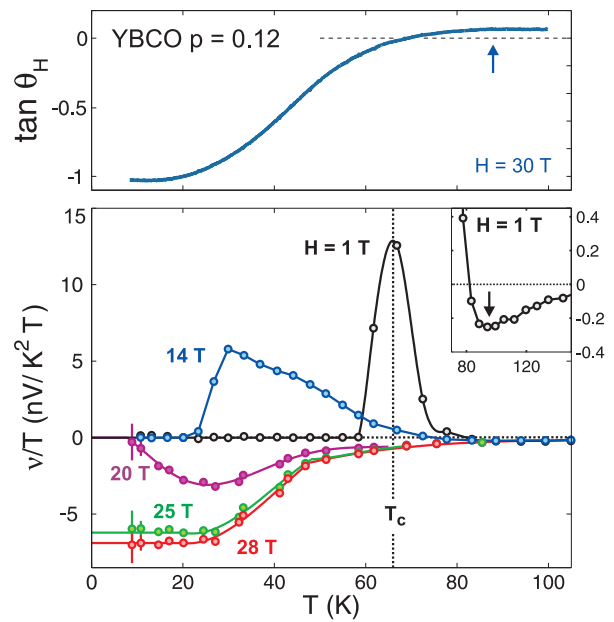}
\caption{ 
Upper panel:
Hall angle $\theta_{H}$ plotted as $\tan \theta_{H} = \rho_{xy}/\rho_{xx}$ vs $T$.
The arrow marks the onset of the drop to large negative values.
Lower panel:
Nernst coefficient $\nu$ plotted as $\nu/T$ vs $T$ for magnetic fields as indicated.
The vertical line marks the zero-field superconducting transition ($T_c$).
{\it Inset:}
Zoom on data at $H=1$~T. The arrow marks the onset of the positive signal due to superconducting fluctuations.
}
\label{fig:fig2}
\end{center}
\end{figure}


The sample was an uncut, unpolished, detwinned crystal of YBCO grown in a non-reactive BaZrO$_3$ crucible from 
high-purity starting materials \cite{Liang2000}. 
The dopant oxygen atoms ($y = 6.67$) were made to order into an ortho-VIII superstructure, yielding a superconducting transition temperature $T_c = 66.0$~K. 
Transport properties were measured via gold evaporated contacts (resistance $<~1~\Omega$), in a six-contact geometry.
The hole concentration (doping) $p = 0.12$ was determined from a relationship between $T_c$ and the $c$-axis lattice constant \cite{Liang2006}.  
The thermal gradient $\Delta T$ was applied along the $a$-axis and the field $H$ along the $c$-axis.
Measurements of the Seebeck and Nernst coefficients, described elsewhere \cite{DaouPRB2009,Cyr-Choiniere2009}, were performed at Sherbrooke up to 15~T and at the GHMFL
in Grenoble up to 28 T. 
 Measurements of the longitudinal ($\rho_{xx}$) and transverse ($\rho_{xy}$) resistivity, described elsewhere \cite{LeBoeuf2007}, were performed at the NHMFL in Tallahassee up to 45~T ($\rho_{xy}$ data were reported in \cite{LeBoeuf2007}).


The Nernst and Seebeck coefficients are plotted as a function of magnetic field in Fig.~1.
At $T>80$, the Seebeck coefficient S is essentially field independent.
At $T<80$, S exhibits a weak field dependence at high field above the vortex solid 
phase (where $\nu = S = 0$),
which we attribute to magnetoresistance \cite{LeBoeuf2007} given that $S$ is the energy derivative of the conductivity \cite{Behnia2004}. 
The Nernst coefficient $\nu/T$ develops a strong positive peak above the 
melting line  due to vortex motion in the vortex liquid phase.
It is followed by a gradual descent to negative values until $\nu(H)$ becomes flat as the field approaches 30~T.
At our lowest temperature, $T = 9$~K, 
$d\nu/dH$ $\to$~0 at $H \simeq 30$~T (inset of Fig.~1).
This shows that the positive vortex contribution to $\nu$ has been suppressed to nearly zero by 28~T, so that the curve of $\nu/T$ vs $T$ at $H = 28$~T shown in Fig.~2 
can be regarded as representative of the normal-state Nernst coefficient of YBCO at $p = 0.12$.
The fact that superconductivity can be suppressed by such a modest field is 
special to $p\simeq 0.12$, where it is weakened by a competing tendency towards stripe order~\cite{Ando2002}. An estimate of the mean-field upper critical field from 
the measured fluctuation magneto-conductivity yields $H_{\rm c2}$(0) = 35~T~\cite{Ando2002} 
-- much smaller than the na\"ive estimate from the Fermi velocity ($\simeq 150$~T).

At low fields, the vortex signal shows up as a large peak centered on $T_c$ (see 1 T data in Fig.~2), which vanishes by 90 K (inset of Fig.~2), 
{\it i.e.} some 25 K above $T_c$, in agreement with previous measurements \cite{Rullier-Albenque2006}. 
The onset of the vortex signal would therefore seem to coincide
   roughly with the onset of diamagnetism, known to occur some 30 K
   above $T_c$ in YBCO near optimal doping \cite{Li2009}.

Having established that 28~T is sufficient to access the normal state down to 9~K, we now examine the normal-state Seebeck coefficient $S(T)$, displayed in Fig.~3.
We see that $S(T)$ undergoes a change of sign at $T \simeq 50$~K, from positive above to negative below.
This is similar to the sign change at $T \simeq 70$~K reported earlier for the Hall coefficient $R_{\rm H}(T)$ \cite{LeBoeuf2007}. 
The fact that both $S$ and $R_{\rm H}$ are negative in the normal state at $T \to 0$ is compelling evidence for an electron-like sheet in the Fermi surface.
That this electron-like sheet dominates over other hole-like portions of the Fermi surface shows that it must have a higher mobility. 
Given that the amplitude of quantum oscillations is exponentially dependent on mobility, it is very likely that this electron-like sheet is the small closed Fermi pocket detected by quantum oscillations in YBCO at a similar doping \cite{Doiron-Leyraud2007,Bangura2008,Jaudet2008}. 
This conjecture is supported by an overall quantitative consistency, as we now show.


\begin{figure}
\begin{center}
\includegraphics[width=0.44\textwidth]{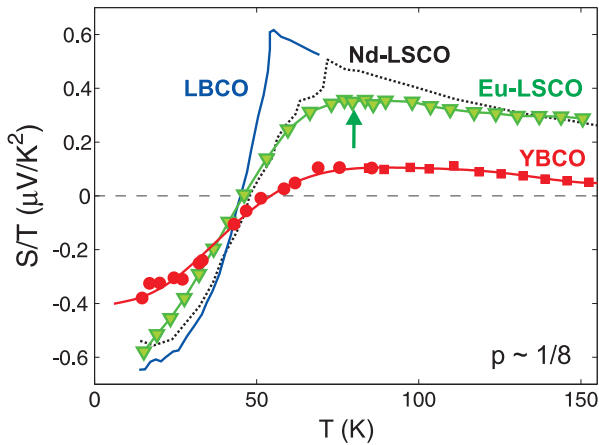}
\caption{
Thermopower of four hole-doped cuprates at $p \simeq 1/8$, plotted as $S/T$ vs $T$: 
YBa$_2$CuO$_{6.67}$ (YBCO, $H=28$~T (circles) and $H=0$ (squares); this work), 
La$_{1.675}$Eu$_{0.2}$Sr$_{0.125}$CuO$_4$ (Eu-LSCO, $H=0$, triangles; this work), 
La$_{1.48}$Nd$_{0.4}$Sr$_{0.12}$CuO$_4$ (Nd-LSCO, $H=0$, dotted line; ref.~\cite{Nakamura1992}), 
and La$_{1.875}$Ba$_{0.125}$CuO$_4$ (LBCO, $H=9$~T, solid line; ref.~\cite{Li2007}). 
The arrow marks the onset of stripe order in Eu-LSCO at $p = 0.125$ measured by X-ray diffraction \cite{Cyr-Choiniere2009,Fink2008}.
}
\label{fig:fig4}
\end{center}
\end{figure}

In a two-band model of electrons ($e$) and holes ($h$)  the two types of carriers will respectively make negative and positive contributions to both $R_H$ and $S$.
It has recently been shown that such a model can account in detail for the field and temperature dependence of $\rho_{xx}$ and $R_H$ in the closely-related material YBa$_2$Cu$_4$O$_8$~\cite{Rourke2009}.
For the measured value of $S$ at $T \to 0$ to come out negative, at $S/T = -~0.4~\mu$V K$^{-2}$ (Fig.~3), we must have $|S_e/T| > 0.4~\mu$V K$^{-2}$, given that the hole-like carriers will contribute a compensating positive term.
From the quantum oscillations (measured on YBCO crystals with slightly lower doping but nearly identical $R_{\rm H}(T)$ \cite{Doiron-Leyraud2007,LeBoeuf2007}), 
we obtain 
$T_{\rm F} = (e \hbar / k_{\rm B}) (F / m^\star) = 410 \pm 20$~K,
in terms of the oscillation frequency $F = 540 \pm 4$~T 
and cyclotron mass $m^\star = 1.76 \pm 0.07~m_0$ \cite{Jaudet2008}, where $m_0$ is the electron mass.  
From Eq.~1, this yields $| S/T | = 1.0~\mu$V K$^{-2}$, which is indeed greater than $0.4~\mu$V K$^{-2}$.
This first consistency check shows that the Fermi pocket measured by quantum oscillations has a sufficiently small Fermi energy to account for the large negative
thermopower at $T \to 0$.

A second consistency check is on the carrier mobility $\mu$. 
The quantum oscillations come from carriers with a mobility $\mu = 0.02~\pm~0.006$~T$^{-1}$ \cite{Jaudet2008}.
We can estimate the transport mobility, from the Hall angle $\theta_{\rm H}$, via 
tan~$\theta_{\rm H} \simeq \mu H$. In Fig.~2, we plot tan~$\theta_{\rm H}$ vs $T$ at $H = 30$~T, which saturates to a value of $-1.0$ below 20 K.
 This yields $\mu \simeq 0.033$~T$^{-1}$, a value consistent with quantum oscillations, if we note that transport mobilities are always somewhat higher since transport is not affected by small-angle scattering whereas quantum oscillations are \cite{Shoenberg}.
 This shows that it is reasonable to attribute the negative sign of $R_{\rm H}$ to the electron-like nature of the Fermi pocket 
 responsible for the quantum oscillations.

The third consistency check is on the Nernst coefficient.
In the $T \to 0$ limit, the magnitude of $\nu/T$ should be approximately equal to $2\mu/3~| S / T |$ (from Eqs. 1 and 2).
In YBCO at $T \to 0$, we find $|\nu/T| \simeq 7$~nV~/~K$^2$~T (Fig.~2) and $2\mu/3~|S/T| \simeq 9$~nV~/~K$^2$~T (using $\mu = 0.033$~T$^{-1}$).
Although it says nothing about the sign of the carriers, this good agreement shows that the large value of the quasiparticle Nernst coefficient in YBCO (comparable in magnitude, but opposite in sign, to the vortex signal at low fields) is due to a combination of small Fermi energy and high mobility, in much the same ratio as would be obtained from the carriers responsible for the quantum oscillations.


A natural explanation for the origin of the electron pocket is a reconstruction of the original large hole-like Fermi surface by some density-wave order that breaks translational symmetry \cite{LeBoeuf2007,Chakravarty2008,Taillefer2009}.  Two kinds of order appear to be likely candidates, both involving a spin-density-wave (SDW).
The only form of order that has been observed unambiguously in YBCO so far is a SDW with wavevector $Q = (0.5\pm \delta, 0.5)$, 
where $\delta \simeq 0.06$, detected recently by neutron diffraction at $p \simeq 0.08$ ($T_c = 35$~K) \cite{Haug2009}.
Calculations show that the Fermi surface reconstruction caused by this type of SDW order would produce an electron pocket of the right size and mass \cite{Harrison2009}. 
The question is whether this SDW order could persist up to $p = 0.12$ once superconductivity is removed. 
This is not inconceivable since the SDW phase in La$_{2-x}$Sr$_x$CuO$_4$ (LSCO) (where $\delta \simeq 0.12$), for example, is known to be extended to higher doping when a magnetic field is applied to suppress superconductivity \cite{Khaykovich2005,Chang2008}.

The second candidate order is the so-called `stripe order', a SDW with $\delta \simeq 1/8$, as found in LSCO or more prominently in three cuprate materials with the low-temperature tetragonal (LTT) structure, namely La$_{2-x-y}$Nd$_y$Sr$_x$CuO$_4$ (Nd-LSCO) \cite{Ichikawa2000}, La$_{2-x-y}$Eu$_y$Sr$_x$CuO$_4$ (Eu-LSCO) \cite{Hucker2007} and La$_{2-x}$Ba$_x$CuO$_4$ (LBCO) \cite{Fujita2004}, where it also involves a charge-density-wave (CDW) with wavevector $Q = (0, 0 \pm 2\delta)$.
Theoretically, the effect of such stripe order on the Fermi surface of a hole-doped cuprate at $p=1/8$ was shown to cause a reconstruction which generically yields an electron pocket \cite{Millis2007,Millis2008}.
Experimentally, stripe order was shown to cause an enhancement of the quasiparticle 
contribution to $\nu/T$ in Nd-LSCO and Eu-LSCO \cite{Cyr-Choiniere2009}.
Note that in the latter materials the enhanced quasiparticle $\nu$ is positive; 
recent calculations suggest the sign of $\nu$ may reflect a difference in $Q$-vector ($\delta = 1/16$ vs 1/8) \cite{Hackl2009}.

In Fig.~3, we show the Seebeck coefficient of LBCO \cite{Li2007}, Nd-LSCO \cite{Nakamura1992} and Eu-LSCO, all at $p \simeq 1/8$. 
$S(T)$ is almost identical in all three materials, with $S/T$ dropping just below the onset of stripe order in each case and crossing to negative values below $T \simeq 50$~K. 
There is no doubt that the change of sign in these materials is associated with stripe order \cite{Cyr-Choiniere2009}.
This strongly suggests that stripe order causes a Fermi surface reconstruction, whose manifestation 
is a change of sign in the Hall and Seebeck coefficients.
Note that in LBCO and Nd-LSCO, the onset of stripe order coincides with the LTT transition, causing the drop in $S/T$ to be sharp, while in Eu-LSCO the LTT transition (at 130~K \cite{Hucker2007}) occurs well above the onset of stripe order (at 80~K \cite{Cyr-Choiniere2009,Fink2008}), producing a smooth drop in $S/T$.     
The behaviour of $S/T$ in YBCO at the same doping is remarkably similar, with a sign change also at $T \simeq 50$~K and a comparable value at $T \to 0$.  
A striking similarity also shows up in 
$R_{\rm H}(T)$, which drops in identical fashion in Eu-LSCO and YBCO \cite{Taillefer2009}.

Given the similarities in $S/T$ and $R_H$~\cite{LeBoeuf2007,Taillefer2009},
we conclude that
an electron Fermi pocket is a common feature of all four hole-doped cuprates near $p = 1/8$. 
Theoretically, this is most easily explained by a reconstruction of the Fermi surface caused by
density-wave order \cite{Millis2008,Hackl2009}. 
Although we cannot exclude exotic orders such as d-density-wave order \cite{Chakravarty2001}, a spin/charge density 
wave appears to be the most likely candidate at $p=1/8$. 
Our findings call for a search for spin/charge density-wave order in YBCO at $p=1/8$
via neutron/X-ray diffraction or NMR/NQR in high magnetic fields.

Upon cooling, the Fermi surface reconstruction in YBCO at $p = 0.12$ begins at 90~K or so, as detected by the drop in $S/T$ and $\tan \theta_{H}$.
(Note that this drop is independent of field (see Fig.~S1b in \cite{LeBoeuf2007}), so that the reconstruction occurs in zero field, at least for $T > T_c$. 
For $T < T_c$, phase competition with superconductivity most likely suppresses the SDW phase \cite{Sachdev2009}.)
Further studies are underway to determine the doping dependence of this onset temperature, and see whether it goes to zero at a critical doping comparable to the quantum critical point for stripe order in Nd-LSCO ({\it i.e.} $p \simeq 0.24$) \cite{DaouPRB2009,Daou2009}.

Our Nernst measurements in YBCO reveal that the quasiparticle contribution in cuprates can be as large as the vortex contribution, on which most of the attention has been focused until now \cite{Rullier-Albenque2006,OngPRB2006}. We expect this quasiparticle contribution, which can be of either sign, to dominate the Nernst signal well above $T_c$, as found in the hole-doped cuprate Eu-LSCO \cite{Cyr-Choiniere2009} and the electron-doped cuprate Pr$_{2-x}$Ce$_x$CuO$_4$ \cite{PCCO}, where quasiparticle and vortex signals have also been disentangled.


We thank J. Flouquet and L. Balicas for their assistance with measurements at the GHMFL and NHMFL, respectively.
J.C. was supported by a Fellowship from the Swiss SNF and 
FQRNT.
Part of this work was supported by Euromagnet under the EU contract RII3-CT-2004-506239.
C.P. and K.B. acknowledge support from the ANR project DELICE.
R. L., D. B., and W. H. acknowledge support from NSERC
L.T. acknowledges support from the Canadian Institute for
Advanced Research, a Canada Research Chair, NSERC, CFI and FQRNT.



\begin{thebibliography}{99}

\bibitem{Norman2005}
M.R. Norman \textit{et al.}, Adv. Phys. \textbf{54}, 715 (2005).

\bibitem{Vignolle2008} 
B. Vignolle \textit{et al.}, Nature \textbf{455}, 952 (2008).

\bibitem{LeBoeuf2007}
D. LeBoeuf \textit{et al.}, Nature \textbf{450}, 533 (2007).


\bibitem{Doiron-Leyraud2007}
N. Doiron-Leyraud \textit{et al.}, Nature \textbf{447}, 565 (2007).


\bibitem{Yelland2008}
E. A. Yelland \textit{et al.}, Phys. Rev. Lett. 100, 047003 (2008).
  
\bibitem{Bangura2008}
A.F. Bangura \textit{et al.}, Phys. Rev. Lett. \textbf{100}, 047004 (2008).

\bibitem{Jaudet2008}
C. Jaudet {\it et al}., Phys. Rev. Lett. {\bf 100}, 187005 (2008).


\bibitem{Chakravarty2008}
S. Chakravarty, Science \textbf{319}, 735 (2008).

\bibitem{Behnia2004}
K. Behnia \textit{et al.}, J. Phys.: Condens. Matter \textbf{16}, 5187 (2004).

\bibitem{Behnia2009}
K. Behnia, J. Phys.: Condens. Matter \textbf{21}, 113101 (2009).

\bibitem{Liang2000}
R. Liang {\it et al}., Physica C {\bf 336}, 57 (2000).

\bibitem{Liang2006}
R. Liang {\it et al}., Phys. Rev. B {\bf 73}, 180505 (2006).

\bibitem{DaouPRB2009}
R. Daou {\it et al}., Phys. Rev. B {\bf 79}, 180505 (R) (2009).

\bibitem{Cyr-Choiniere2009}
O. Cyr-Choini\`ere \textit{et al.}, Nature \textbf{458}, 743 (2009).

\bibitem{Ando2002}
Y. Ando and K. Segawa, Phys. Rev. Lett. {\bf 88}, 167005 (2002).

\bibitem{Rullier-Albenque2006}
F. Rullier-Albenque {\it et al}., Phys. Rev. Lett. {\bf 96}, 067002 (2006).


\bibitem{Li2009}
L. Li {\it et al}., arXiv:0906.1823. 

\bibitem{Rourke2009}
P. M. C. Rourke {\it et al}., arXiv:0912.0175.

\bibitem{Shoenberg}
D. Shoenberg, {\it Magnetic oscillations in metals} (CUP, Cambridge, 1984).

\bibitem{Taillefer2009}
L. Taillefer, J. Phys.: Condens. Matter {\bf 21}, 164212 (2009).

\bibitem{Haug2009}
D. Haug {\it et al}., Phys. Rev. Lett. {\bf 103}, 017001 (2009).

\bibitem{Harrison2009}
N. Harrison, Phys. Rev. Lett. {\bf 102}, 206405 (2009).

\bibitem{Khaykovich2005} 
B. Khaykovich {\it et al}., Phys. Rev. B {\bf 71}, 220508(R) (2005).

\bibitem{Chang2008} 
J. Chang {\it et al}., Phys. Rev. B {\bf 78}, 104525 (2008);
{\it ibid.}, Phys. Rev. Lett. {\bf 102}, 177006 (2009).

\bibitem{Ichikawa2000}
N. Ichikawa {\it et al}., Phys. Rev. Lett. {\bf 85}, 1738 (2000).

\bibitem{Hucker2007}
M. Hucker {\it et al.}, Physica C {\bf 460}-{\bf 462}, 170 (2007).

\bibitem{Fujita2004} 
M. Fujita {\it et al}., Phys. Rev. B {\bf 70}, 104517 (2004).

\bibitem{Millis2007}
A.J. Millis and M.R. Norman, Phys. Rev. B \textbf{76}, 220503 (2007).

\bibitem{Millis2008}
J. Lin and A.J. Millis, Phys. Rev. B \textbf{78}, 115108 (2008).

\bibitem{Hackl2009}
A. Hackl, M. Vojta and S. Sachdev, Phys. Rev. B \textbf{81}, 045102 (2010).

\bibitem{Li2007}
Q. Li \textit{et al.}, Phys. Rev. Lett. \textbf{99}, 067001 (2007).

\bibitem{Nakamura1992}
Y. Nakamura and S. Uchida, Phys. Rev. B \textbf{46}, 5841 (1992).

\bibitem{Fink2008}
J. Fink {\it et al}., Phys. Rev. B {\bf 79}, 100502(R) (2009). 

\bibitem{Sachdev2009}
S. Sachdev, arXiv:0907.0008.

\bibitem{Chakravarty2001}
S. Chakravarty {\it et al}.,  Phys. Rev. B \textbf{63}, 094503 (2001).

\bibitem{Daou2009}
R. Daou {\it et al}., Nature Phys. {\bf 5}, 31 (2009).

\bibitem{OngPRB2006}
Y. Wang {\it et al}.,  Phys. Rev. B \textbf{73}, 024510 (2006).

\bibitem{PCCO}
P. Li and R.L. Greene, Phys. Rev. B \textbf{76}, 174512 (2007).




\end{thebibliography}
\end{document}